\documentclass[aps, prd, reprint, superscriptaddress, nofootinbib, 10pt]{revtex4-2}

\usepackage{blindtext}
\usepackage{graphicx}
 
\usepackage{comment}
\usepackage{amsmath}
\usepackage{lineno}
\usepackage{hyperref}

\begin{document}
\title{First Nuclear Ultra-Heavy Dark Matter Search in Argon Time Projection Chambers with the DarkSide-50 Experiment}


\author{P.~Agnes}\affiliation{Gran Sasso Science Institute, L'Aquila 67100, Italy}\affiliation{INFN Laboratori Nazionali del Gran Sasso, Assergi (AQ) 67100, Italy}
\author{I.~F.~Albuquerque}\affiliation{Instituto de F\'isica, Universidade de S\~ao Paulo, S\~ao Paulo 05508-090, Brazil}
\author{T.~Alexander}\affiliation{Pacific Northwest National Laboratory, Richland, WA 99352, USA}
\author{A.~K.~Alton}\affiliation{Physics Department, Augustana University, Sioux Falls, SD 57197, USA}
\author{M.~Ave}\affiliation{Instituto de F\'isica, Universidade de S\~ao Paulo, S\~ao Paulo 05508-090, Brazil}
\author{H.~O.~Back}\affiliation{Pacific Northwest National Laboratory, Richland, WA 99352, USA}
\author{G.~Batignani}\affiliation{Physics Department, Universit\`a degli Studi di Pisa, Pisa 56127, Italy}\affiliation{INFN Pisa, Pisa 56127, Italy}
\author{K.~Biery}\affiliation{Fermi National Accelerator Laboratory, Batavia, IL 60510, USA}
\author{V.~Bocci}\affiliation{INFN Sezione di Roma, Roma 00185, Italy}
\author{W.~M.~Bonivento}\affiliation{INFN Cagliari, Cagliari 09042, Italy}
% ...

\begin{abstract}
We report the first search for nuclear ultra-heavy dark matter (UHDM) in a dual-phase liquid argon time projection chamber using the DarkSide-50 experiment. Unlike conventional weakly interacting massive particles (WIMPs), a nuclear UHDM model may be composed of many dark nucleons and scatter numerous times while passing through the detector. Accounting for energy loss through the Earth's overburden, we apply selection criteria optimized for multi-scatter event topologies using the 532-day low-radiation campaign of the DarkSide-50 detector. Excluded limits on the UHDM-nucleon scattering cross section for dark nucleon masses of $m_\chi = 10, 50, 100, 500 \, \mathrm{GeV/c^2}$ are presented.
  
\end{abstract}

\maketitle

\section{Introduction}
\label{sec:Introduction}
Evidence for dark matter is well documented at different cosmological scales. Velocity distribution of galaxies \cite{Faber}, galaxy cluster mergers \cite{Okamoto}, cosmic microwave background observations (CMB) from Planck \cite{Planck}, and simulations and observations of the large-scale structure of the universe \cite{Spergel}, all provide strong supporting evidence for the existence of cold dark matter. Among the many different dark matter models, weakly interacting massive particles (WIMPs) are well-established within the scientific community, and may provide a detection channel via low-energy elastic scattering \cite{Goodman}. In this work, we extend the DarkSide-50 dark matter search to nuclear ultra-heavy dark matter (UHDM), investigating composite dark nuclei that would produce distinctive multiple-scatter signatures in the detector. 

Hosted at the Laboratori Nazionali del Gran Sasso (LNGS), Italy, DarkSide-50 is a dual phase time projection champer (TPC) filled with purified liquid argon (LAr) \cite{DarkSideFirst,DS50Argon}. The experiment was primarily designed for a background-free search of WIMPs, obtaining a 90\% Confidence Level (CL) on the spin-independent (SI) dark matter-nucleon scattering cross section at $1.14\times 10^{-44}\,\mathrm{cm^2}$ ($3.43\times  10^{-43}\,\mathrm{cm^2}$) 
for a WIMP mass of $100 \, \mathrm{GeV/c^2}$ ($1\, \mathrm{TeV/c^2}$) \cite{DarkSide-50-532}. The main active component of DarkSide-50 was a LAr TPC. The TPC was a cylindrical detector with a radius of 18 cm and a  height of 36 cm, and was viewed by two arrays of 19 Hamamatsu R11065 photo multiplier tubes (PMT). The detector was surrounded by a 4 m diameter sphere filled with liquid scintillator acting as a neutron veto, and a 10 m tall and 11 m in diameter water Cherenkov detector which served as a muon detector \cite{DarkSideVeto}. Two signals provide the necessary information for the identification of a potential candidate: a prompt scintillation signal (S1) produced by particles elastically scattering off LAr nuclei, and a delayed electroluminescence signal (S2) produced by extracting the ionization electrons from the bulk LAr into a gaseous argon (GAr) pocket at the top of the TPC via the application of an electric field.

\begin{figure}[htpb]
    \centering
    \includegraphics[width=8.5cm]{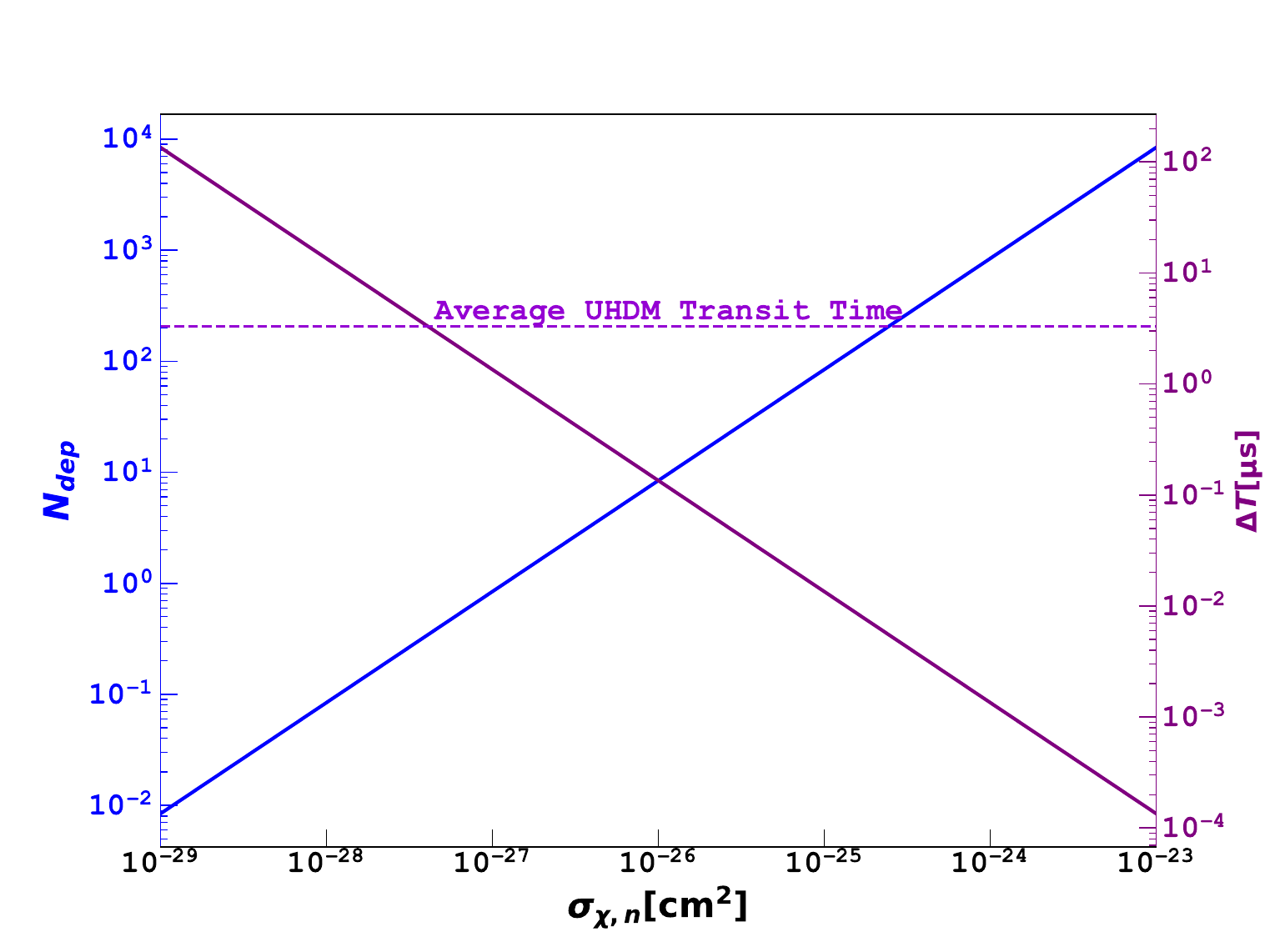}
    \setlength{\abovecaptionskip}{0pt}
    \caption{Average number of energy deposits for UHDM of different $\sigma_{\chi,n}$.The left-hand (blue) axis shows the number of energy deposits, and the right-hand (purple) shows the time between energy deposits. The dashed line represents the average UHDM transit time.}
    \label{fig:ndeps}
\end{figure}

Beyond the traditional WIMP, UHDM represents an alternative model which has been searched in other experiments \cite{DEAPFirst,XenonProjected,Quartz,Lux}. In models where the dark sector lacks a long-range force analogous to electromagnetism, dark nucleosynthesis can proceed unconstrained, potentially forming composite dark matter nuclei containing up to $10^{20}$ constituents \cite{Hardy, Coskuner, Khlopov2}. UHDM particles would interact with ordinary matter through elastic scattering off nuclei, similar to WIMPs but with distinctive signatures due to their composite structure \cite{Butcher}. The most distinctive aspect of such a search is the multi-scatter nature of UHDM particles on their pass through the TPC. Figure \ref{fig:ndeps} shows the average number of energy deposits ($N_{\mathrm{dep}}$) expected for a UHDM pass through the TPC. The dashed line indicates the average UHDM transit time, $\tilde{t}_{\mathrm{cross}} = 3.3 \, \mathrm{\mu s}$, assuming an average UHDM velocity of $v_{\mathrm{UHDM}}=220 \,\mathrm{km/sec}$. 
This work explores the sensitivity of DarkSide-50 to UHDM for different dark nucleon masses. Unlike prior searches, which treat UHDM as a point-like particle and rely on the standard $A^2$ coherent enhancement scaling, this analysis introduces $m_{\chi}$ as an explicit parameter of the model, introducing the UHDM form factor $F_\chi(q, R_\chi)$ and producing signal hypotheses that are not directly comparable to those of other experiments.

\section{Nuclear Ultra-Heavy Dark Matter}
\label{sec:HDM}
The formation of nuclear ultra-heavy dark matter (UHDM), dark matter nuclei which are composed of many dark nucleons, may be explained by the absence of the analogue long-range electromagnetic interaction in the dark sector \cite{Hardy}, which allows for UHDM nuclei of up to $10^{20}$ constituents \cite{Coskuner} to be formed via Big-Bang Dark Nucleosynthesis (BBDN) \cite{Hardy}. Similarly to WIMPs, UHDM can also interact with Standard Model (SM) particles by elastically scattering off nuclei, providing a detection mechanism.

The differential dark matter-SM nucleus scattering cross section is described by:
\begin{equation}
        \frac{d\sigma_{\chi,N}}{dE_R} = \frac{m_N }{2\mu^2_{\chi,N}v^2}\sigma_{\chi,n}\left(\frac{\mu_{\chi,N}}{\mu_{\chi,n}}\right)^{2}A^2F^{2}_{\chi}(q)F^{2}_{N}(q).
\label{eq:dcs}
\end{equation}
Here, $\sigma_{\chi,N}$ is the dark matter- SM nucleus cross section,  $\sigma_{\chi,n}$ is the dark matter nucleus-nucleon cross section, $\mu_{\chi,n}$ is the dark matter nucleus-nucleon reduced mass, $\mu_{\chi,N}$ is the DM-target reduced mass, $A$ is the nucleus atomic mass number, $m_N$ is the mass of the target nuclei, $M_{\chi}$ is the mass of the UHDM nucleus, $E_R$ is the recoil energy, and $F_{\chi}$ and $F_{N}$ are the dark matter and SM nucleus form factors.
$F_{N}$ is the Helm form factor \cite{Helm}, while the dark matter form factor, $F_{\chi}$, is described by \cite{Butcher}:
\begin{equation}
    F_{\chi}(q,R_{\chi}) = \frac{3j_1(qR_{\chi})}{qR_{\chi}}
\label{eq:FF}
\end{equation}
where $j_1$ is a Bessel function of the first kind and $q$ is the momentum transfer. Equation \ref{eq:FF} assumes the dark matter nuclei have a uniform density which is modeled by a spherical top hat function \cite{Butcher}. Consequently, the introduction of the finite size of the dark matter nucleus suppresses the energy of the recoiling nuclei, favoring low-energy scatters. 

Assuming a dark nucleon forces equivalent to the SM, the radius of the UHDM nucleus may be characterized by \cite{Coskuner}:
\begin{equation}
    R_{\chi} = \left(\frac{9\pi M_{\chi}}{4m^4_{\chi}}\right)^{1/3}
    \label{eq:radius}
\end{equation}
where $m_{\chi}$ is the mass of the dark nucleon (constituent). In this analysis, $\sigma_{\chi,n}$ is considered for values up to the geometrical cross section, $\sigma_{\mathrm{geo}}=4\pi R_{\chi}^2$, as larger values are unphysical for contact interactions. 

\begin{figure}[h]
    \centering
    \includegraphics[width=8.5cm]{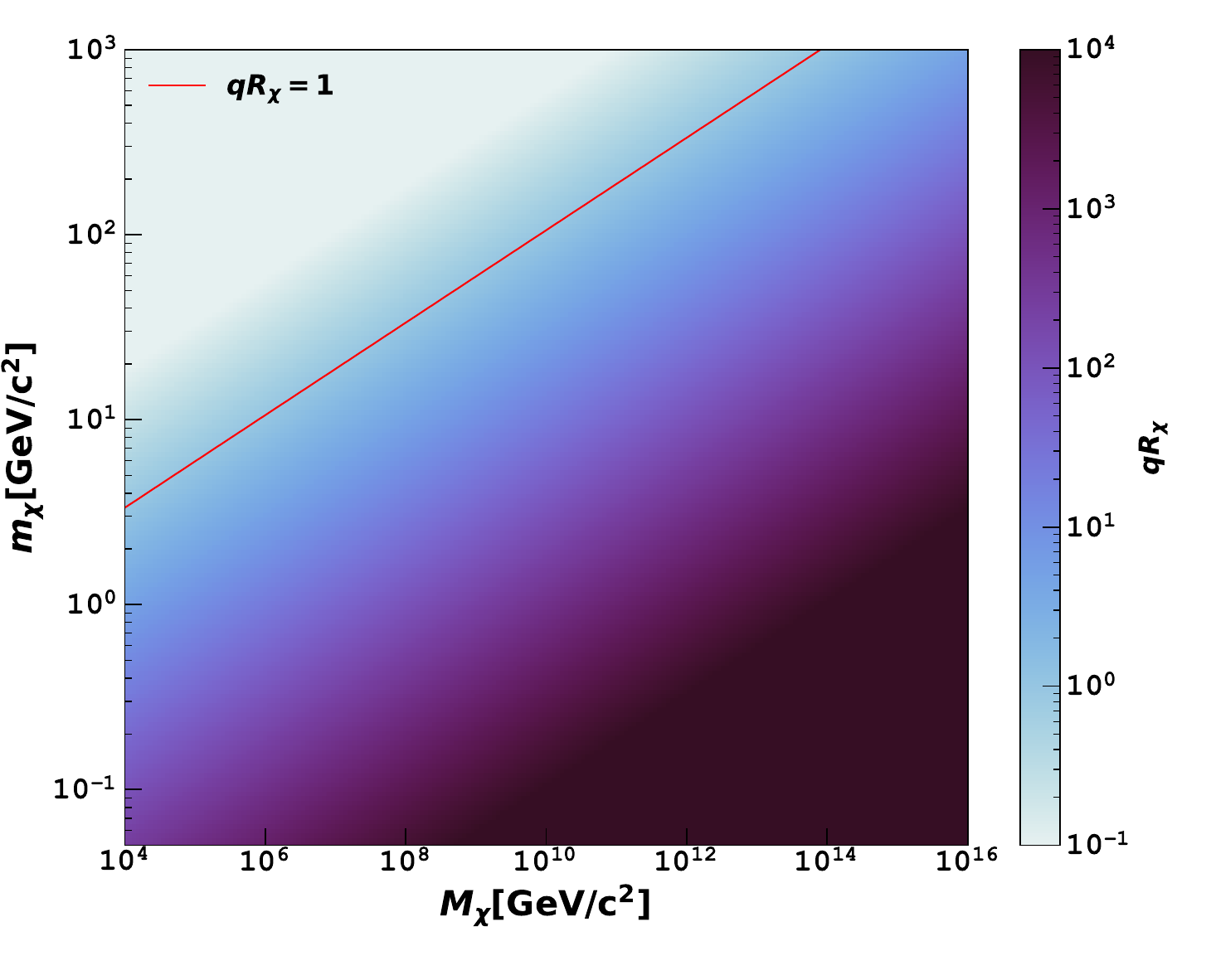}
    \setlength{\abovecaptionskip}{0pt}
    \caption{The product $qR_{\chi}$ as a function of $M_{\chi}$ and $m_{\chi}$. The red line shows where $qR_{\chi}$ = 1; at higher values, coherent scatter tends to be suppressed.}
    \label{fig:Mxmx_left}
\end{figure}

The product $qR_{\chi}$ may be employed to estimate the effect of the form factors: at values of $q_{\mathrm{max}}R_{\chi}\lesssim 1$,  where $q_{\mathrm{max}}=\sqrt{2m_NE^{\mathrm{max}}_R}$ and $E^{\mathrm{max}}_R$ is the maximum kinematically allowed recoil energy determined by the incoming UHDM velocity, the UHDM nucleus size is less than the inverse of the momentum transfer and little form factor suppression occurs \cite{Coskuner}. In this case, the dark matter nucleus can be treated as a point-like particle and coherence enhancement ($A^4$) occurs. This regime is best probed by high-exposure detectors. For $q_{\mathrm{max}}R_{\chi}\gtrsim 1\gtrsim q_{\mathrm{min}}R_{\chi}$, where $q_{\mathrm{min}}=\sqrt{2m_NE^{\mathrm{thr}}_R}$ and $E^{\mathrm{thr}}_R$ is the minimum recoil energy required to produce a detectable signal in the detector, there is intermediate form factor suppression, and coherent enhancement only occurs for a fraction of the phase space. It is in this region where the search presented in this manuscript is conducted. Figure \ref{fig:Mxmx_left} shows the dependence of the $qR_{\chi}$ product as a function of $M_{\chi}$ and $m_{\chi}$. The effects of nuclear form factors become significant for  values of $qR_{\chi}>1$, decreasing the probability of high energy nuclear recoils. 

The search for nuclear dark matter introduces the dark nucleon mass as an additional parameter that needs to be either fixed or evaluated when conducting the search. 
Here, the dark nucleon mass was set to $m_\chi = 10, 50, 100, 500 \, \mathrm{GeV/c^2}$ to evaluate the effect of different dark nucleon masses on the experiment's sensitivity. 

\begin{figure}[h!]
    \centering
    \includegraphics[width=8.5cm]{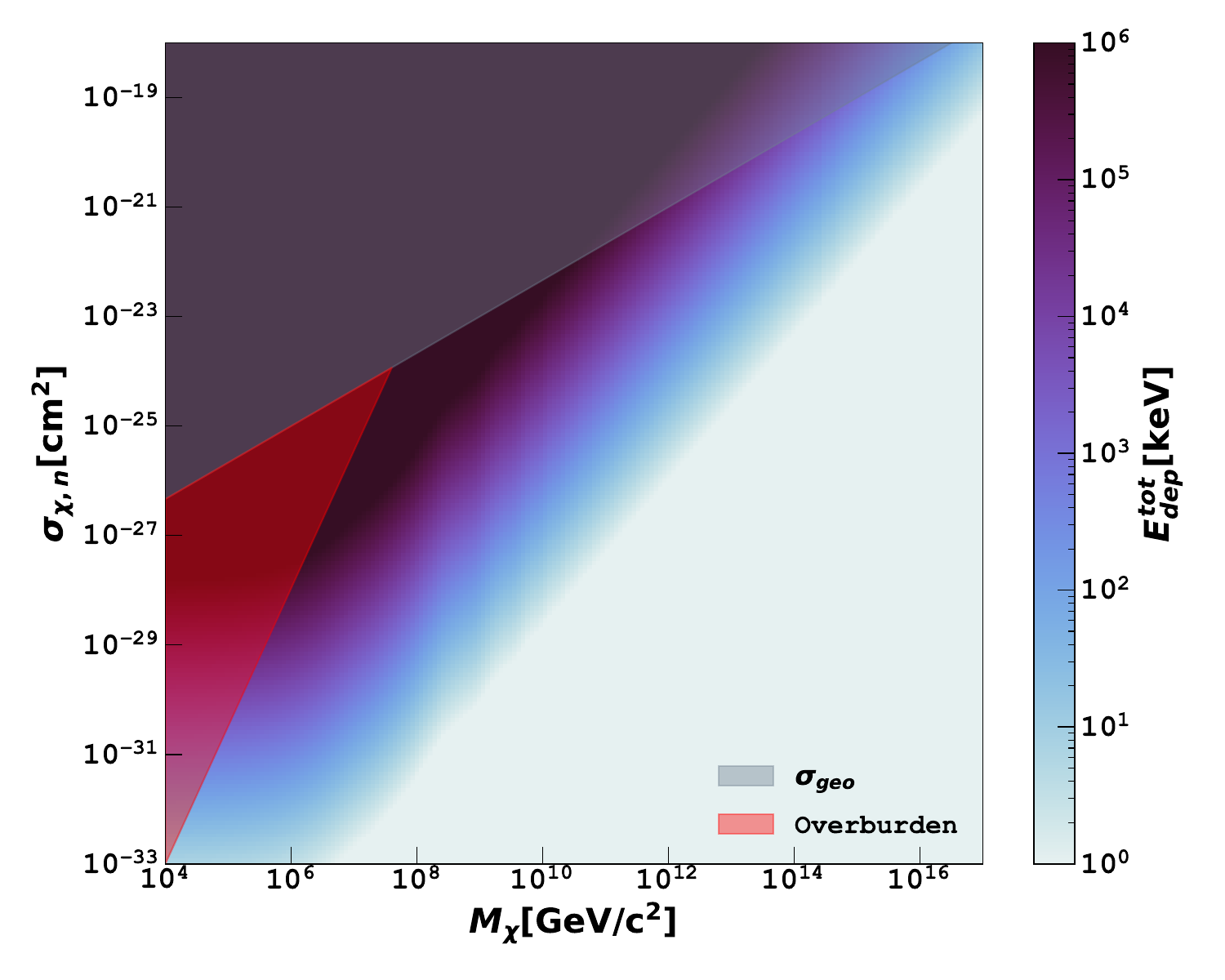}
    \setlength{\abovecaptionskip}{0pt}
    \caption{Energy deposited in the parameter space region of interest, for dark nucleon mass $m_{\chi} = 10 \mathrm{\,GeV/c^2}$. The red region estimates undetectable parameter space due to overburden effects as predicted in \cite{Coskuner}. The gray region delimits area where the scattering cross section $\sigma_{\chi,n}$ is larger than the geometrical cross section $\sigma_{\mathrm{geo}}$, which depends on the dark nucleon mass $m_{\chi}$.}
    \label{fig:pspace}
\end{figure}

The UHDM signals in the detector may be obtained by evaluating the total energy deposited for different UHDM particles in one pass through the TPC. This is illustrated in figure \ref{fig:pspace}, where ($E_{\mathrm{dep}}^{\mathrm{tot}}$ is plotted for a dark nucleon mass of $m_{\chi}=10 \, \mathrm{GeV/c^2}$, and assuming an average UHDM velocity of $v_{\mathrm{UHDM}}=220 \,\mathrm{km/sec}$. 

\section{Overburden}
\label{sec:Overburden}

To accurately calculate the expected signal rate, we account for how Earth's overburden affects UHDM trajectories before reaching the DarkSide-50 detector. This task is accomplished with the Verne toolkit \cite{Kavanagh}, a package that has been employed to conduct other dark matter studies \cite{KavanaghSource}, and that relies on the continuous formalism approach to calculate the energy loss per collision as the UHDM particle travels through Earth. The underlying assumption that makes the continuous formalism suitable is the mass difference between a UHDM and the scattering nucleus ($m_A$), that being $(M_{\chi}\gg m_{A})$. A detailed description of the tool with a discussion of the continuous formalism approach can be found in the original toolkit manuscript \cite{KavanaghSource}. The kinetic energy loss in this path may be written as: 
\begin{equation}
    \frac{d\langle E_{\chi} \rangle}{dx}=-\sum_i n_i(\vec{r})\langle E_{R} \rangle \sigma_{\chi,i}(v),
    \label{eq:kinetic_loss}
\end{equation}
where $n_i(\vec{r})$ is the number density of nuclei species $i$ at position $\vec{r}$ in the detector rest frame, $\sigma_{\chi,i}$ is the DM-$i$ species nucleus scattering cross-section, and $v$ is the DM speed in the detector rest frame. 

\begin{figure}[h]
    \centering
    \includegraphics[width=8.5cm]{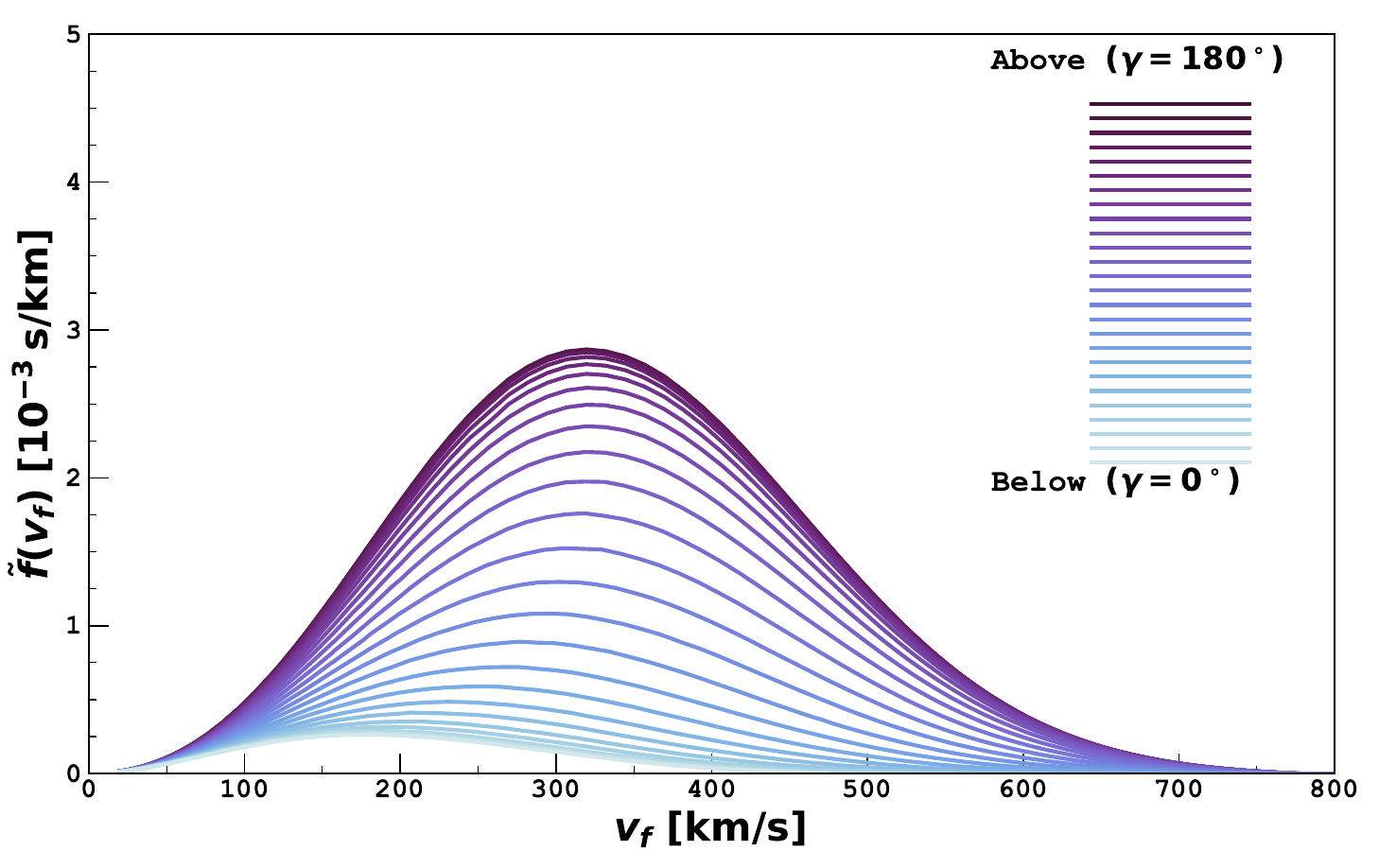}
    \setlength{\abovecaptionskip}{0pt}
    \caption{Velocity distribution in the detector rest frame for a UHDM model of $M_{\chi}=10^{7}\,\mathrm{GeV/c^2}$ and $\sigma_{\chi,n}=10^{-27}\,\mathrm{cm^2}$ for different incoming directions with respect to the detector's vertical axis ($\gamma)$. Darker colors represent particles traveling the shortest path to the detector, while lighter colors represent particles traveling through most of Earth's overburden before reaching the detector.}
    \label{fig:veldist}
\end{figure}

An example of different speed distributions, $\tilde{f}(v_f)$, for a UHDM model of $M_{\chi}=10^{7}\, \mathrm{GeV/c^2}$ and $\sigma_{\chi,n}=10^{-27} \, \mathrm{cm^2}$ for different incoming directions with respect to the detector's vertical is shown in figure \ref{fig:veldist}. The effects of the overburden are most noticeable for model masses below $\sim M_{\chi}=10^{10} \, \mathrm{GeV/cm^2}$, while models of larger masses in the region of interest are unaffected by the overburden. It's also important to introduce the saturated overburden ($\sigma_s$), which occurs when the DM scattering cross-section becomes large enough that dark matter scatters with every nucleus or electron along its path through the overburden material \cite{SaturatedOverburden}. The effects of the water Cherenkov muon veto and liquid scintillator neutron veto on the DM velocity distribution were estimated and considered negligible. 

\section{Detector Response}
\label{sec:response}

\subsection{LAr response}
\label{ssec:LarResponse}
For the cross section values of the considered model, a UHDM nucleus is expected to elastically scatter off LAr nuclei multiple times while passing through the TPC, producing S1 and S2 signals. Several different signal combinations may arise: a series of S1 followed by a series of S2 signals, a series of S1 followed by a series of merged S2 signals, merged S1 followed by merged S2 signals, and a series of unmerged or merged S2 dictated by the lack of S1 signals due to form factor suppression. This last type of event should be unexpected and may also be prevented via fiducialization cut. Given that the dark matter speed average is around $v_0\sim220$ km/sec (in the galactic rest frame), a UHDM nucleus is expected to cross the TPC roughly in $\mathcal{O}(1)\,\mu \mathrm{s}$ timescale, and S2 signals will take $\mathcal{O}(10)\, \mathrm{\mu s}$ to $\mathcal{O}(100)\,\mathrm{\mu s}$ at 200 V/cm nominal drift field. The target signature of this search is a merged S1 pulse, while the charge of the S2 signal is not used as part of the analysis in the search for UHDM. 

From figure \ref{fig:pspace}, UHDM nuclei in the parameter space region of interest will deposit roughly $\mathcal{O}(10)$ keV to $\mathcal{O}(10^3)$ keV. Given this, S1 is employed as the variable of interest. S2 signals are not used in this analysis because the electron drift times in the TPC are comparable to or exceed the transit time of UHDM particles traversing the detector, leading to having to perform model dependent recombination corrections for S2 estimates.  

The LAr scintillation response is modeled to estimate S1. For electronic recoils (ER) \cite{Aris,Darkside-ArSim}:
\begin{equation}
    S1_{ER} = g_1(N_{ex}+N_{i}\times R(E_R,F))
    \label{eq:S1ER}
\end{equation}
where $g_1$ is the collection efficiency of photons, $N_{ex}$ is the number of excitons, $N_i$ is the number of electron-ion pairs, and $R(E_R,F)$ is the recombination probability for an ER of $E_R$ at drift field $F$. The recombination probability for this model is taken from \cite{Darkside-ArSim}.

For nuclear recoils (NR), S1 is taken to be \cite{Aris,Darkside-ArSim}:
\begin{equation}
    S1_{NR} = \frac{g_1}{W} F_{q}(\mathcal{L}_{eff}(E_R) \times E_{R})
    \label{eq:S1NR}
\end{equation}
where $W$ is the effective work function, $F_q(E_R)$ is the NR electric field quenching, and $L_{eff}(E_R)$ is the NR quenching model, both taken from ARIS data, which was fitted against an extended version of the Birks’ formula for organic scintillators \cite{Aris}. 

To estimate whether successive energy deposits in the TPC due to UHDM interactions will lead to a series of unmerged or merged (several signal pulses overlapping in time) S1 signals, a toy peak-finder algorithm that mimics the algorithm in DarkSide-50 was employed to investigate the shape of S1 pulses. Based on the average transit time in figure \ref{fig:ndeps}, it's expected that successive energy deposits merge into a single S1 signal. This was confirmed with a toy simulation modeling UHDM events crossing the TPC via Monte Carlo, converting photo-electron (PE) arrival times to waveforms using measured single-PE response functions, with peaks identified in summed waveforms using a 2 PE threshold. No events had more than one S1 pulse. The average pulse length  was found to be $\bar{t}_{\mathrm{pulse}}=3.9 \pm 2.3 \ \mu s$ where the uncertainty represents one standard deviation of the computed pulse length distribution.

\begin{figure}[htpb]
    \centering
    \includegraphics[width=8.5cm]{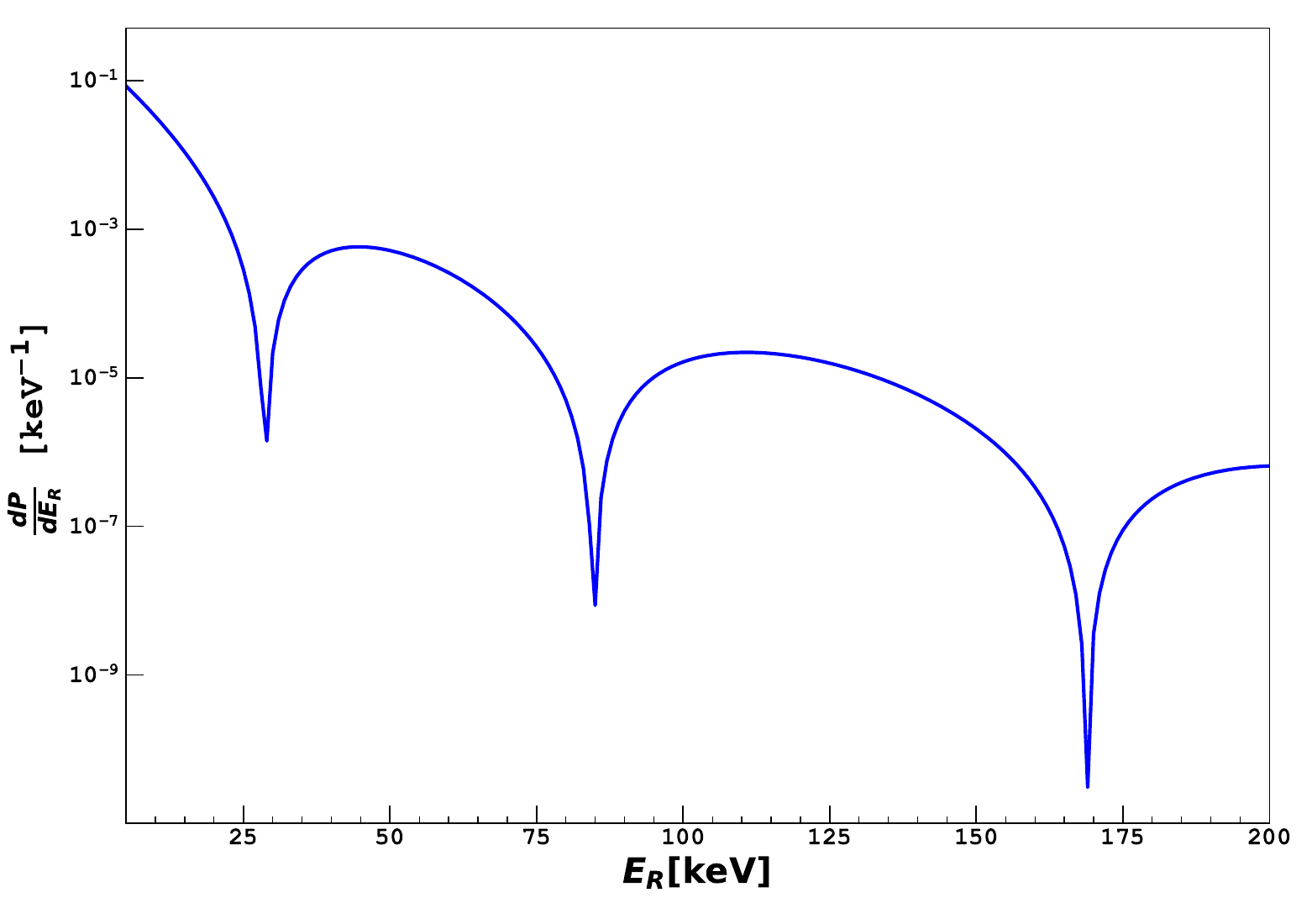}
    \setlength{\abovecaptionskip}{0pt}
    \caption{Differential probability per unit recoil energy for $m_{\chi}=10 \,\mathrm{GeV/c^2}$, $M_{\chi} = 10^7 \,\mathrm{GeV/c^2}$, and $\sigma_{\chi,n}=10^{-27} \, \mathrm{cm^2}$.}
    \label{fig:rate}
\end{figure}

The recoil energy spectrum of a nuclear UHDM particle of $m_{\chi}=10 \,\mathrm{GeV/c^2}$, $M_{\chi} = 10^7 \, \mathrm{GeV/c^2}$, and $\sigma_{\chi,n}=10^{-27} \, \mathrm{cm^2}$ is presented in figure \ref{fig:rate}. Since this search is conducted at cross sections where all UHDM particles are expected to scatter  when passing through the TPC, the number of events in live time $T$ is estimated by \cite{DEAPFirst}:
\begin{equation}
    \label{eq:UHDMrate}
    N = T \frac{\rho_\chi}{m_{\chi}} \int |v|f(\vec{v}) d^3\vec{v}  dA 
\end{equation}
where $A$ is the detector's surface area. 

Since $M_{\chi}\gg m_{A}$, events are modeled as a colinear track of energy deposits passing through the TPC. This is first done by sampling a recoil of energy $E_R$ from the distribution in figure \ref{fig:rate}, where the distinct valleys are the result of the form factor suppression. The angle of incidence of the incoming UHDM nucleus with respect to the detector vertical ($\gamma$) is obtained by sampling from the speed distributions that could have produced an energy deposit $E_R$, where $\theta$ is biased towards values where the overburden is less significant. With $\theta$ defined, the number of energy deposits $(N_{dep})$ within a TPC pass is estimated based on $\sigma_{\chi,n}$ and the length of the track through the detector. Finally, S1 is evaluated by using the sum of all energy deposits ($E^{\mathrm{tot}}_R$) through the TPC as described in section \ref{ssec:LarResponse}. 

\subsection{Background model and data selection} 
\label{ss:bkg}
The data employed for this search is the 532-day underground Ar (UAr) dataset, where the most basic data quality cuts are applied \cite{DarkSide-50-532}. These require that all channels are present, that the baseline be found, and that the event occurs at least 1.21 ms after the previous event to remove events triggered on an S2 where the corresponding S1 was in the inhibit window. The analysis cuts applied progressively and number of survived events are presented in table  \ref{tab:cuts}. The energy region of interest (ROI) is defined to be 100--8000 PE in S1. The low-end boundary (100 PE) 
is defined such that events with total energy deposition of O(10) keV (see figure \ref{fig:pspace}) with several energy deposits will be accepted. The upper bound of the ROI is defined such that most UHDM with total energy depositions up to ~ $\mathcal{O}(1)$ MeV are accepted. At larger recoil energies, the form factor is significant, requiring larger detector volumes. The $t_\text{drift}$ variable is defined as the time difference S1 pulse start and the start of the S2 pulse. To avoid highly interacting events traveling from the cathode towards the anode to produce merged S1 and S2 signals, events with the last energy deposit within 10 $\mu$s of the anode at 200 V/cm drift field are rejected. 

\begin{figure}[htpb]
    \centering
    \includegraphics[width=8.5cm]{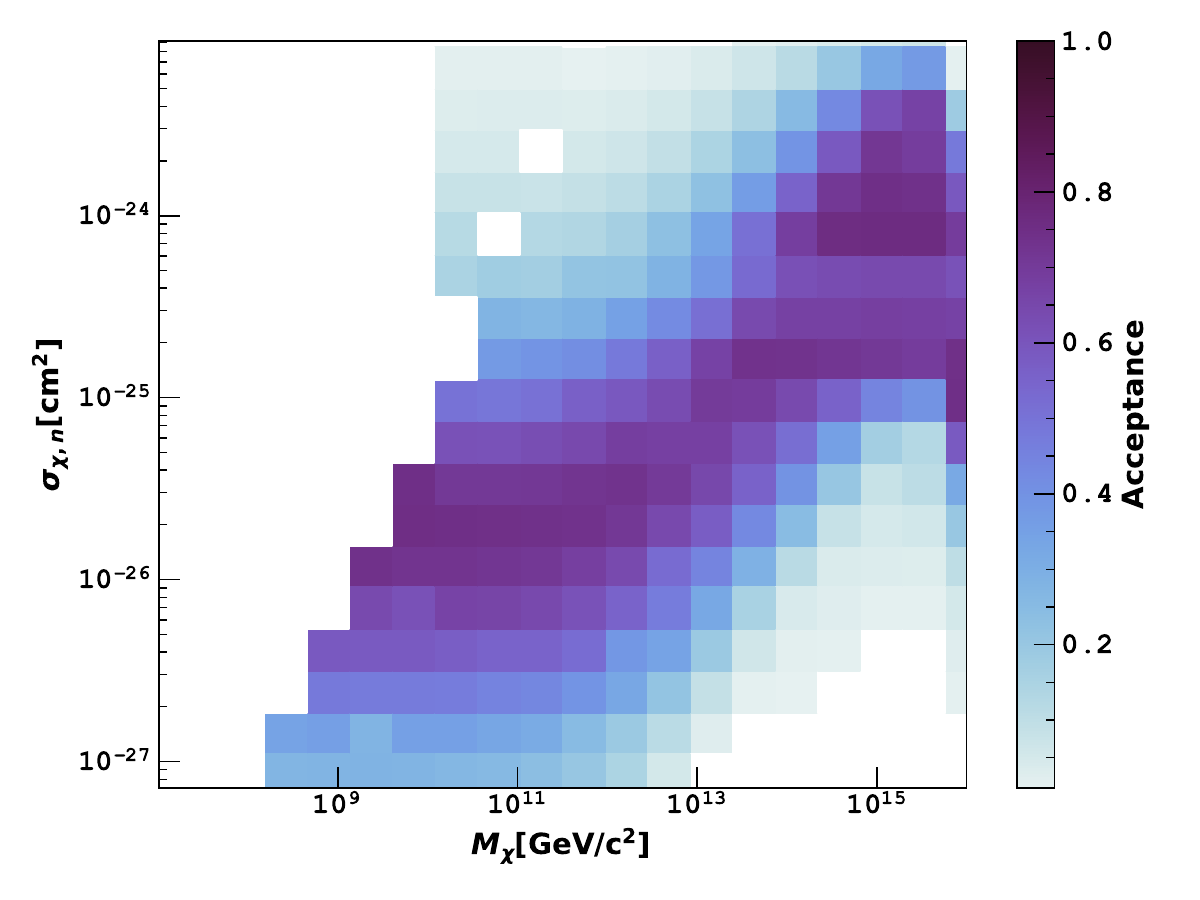}
    \setlength{\abovecaptionskip}{0pt}
    \caption{Fraction of UHDM surviving cuts as a function of $M_{\chi}$ and $\sigma_{\chi,n}$ space for $m_{\chi}=500 \mathrm{\, GeV/c^2}$.}
    \label{fig:acceptance}
\end{figure}

An example of an acceptance (the number of surviving events after the analysis cuts listed in Table \ref{tab:cuts})  distribution in $M_{\chi}$ vs $\sigma_{\chi,n}$ space is shown in figure \ref{fig:acceptance}.
The decreased acceptance around the edges of the band is due to the energy deposited falling above (below) the energy region of interest of 100-8000 PE. The center of this band is the region where most of the energy deposits pass the analysis cuts, with the exception of the $\mathrm{t_{drift}}$ cut removing roughly 15\%-20\% of events that enter the TPC from the top. About  10\% of events are removed due to the imposed condition of not having energy deposits close to the anode to avoid S1 and S2 mergings. The left-hand cut off is due to the UHDM nuclei having an interaction cross section above the geometrical cross section. At very high masses and cross sections, there is an increase in the acceptance due to form factor suppression reducing the amount of energy deposited by UHDM with high interactions. 
\begin{table}[h!]
\centering
\begin{tabular}{l|c}
\hline
\hline
\textbf{Cut} & \boldmath$N_{\text{events}}$ \\ 
\hline
$t_{\text{drift}} > 10\ \mu\text{s}$ & $2.695 \times 10^7$ \\
$t_{\text{drift}} > 10\ \mu\text{s} \, \& \, \text{S1} > 100$~PE & $2.648 \times 10^7$ \\
$t_{\text{drift}} > 10\ \mu\text{s} \, \& \ 100 < \text{S1} < 8000$~PE & $2.351 \times 10^7$ \\ 
\hline
\end{tabular}
\caption{Data cuts and the corresponding remaining number of events after each applied cut. All values include the remaining events after the data quality cuts outlined in the text.}
\label{tab:cuts}
\end{table}

The dominant background in this search is from radiogenic sources. These backgrounds were modeled by Monte-Carlo simulations of the energy depositions originating from the detector's components \cite{DS50Argon}. From these depositions, the S1 distributions may be produced according to the detector response described in section \ref{ssec:LarResponse}. Each contribution is then scaled according to the measured radioactive activities in DarkSide-50 \cite{DarkSide-50-532}.

\subsection{UHDM Energy Spectrum}
\begin{figure}[htpb]
    \centering    
    \includegraphics[width=8.5cm]{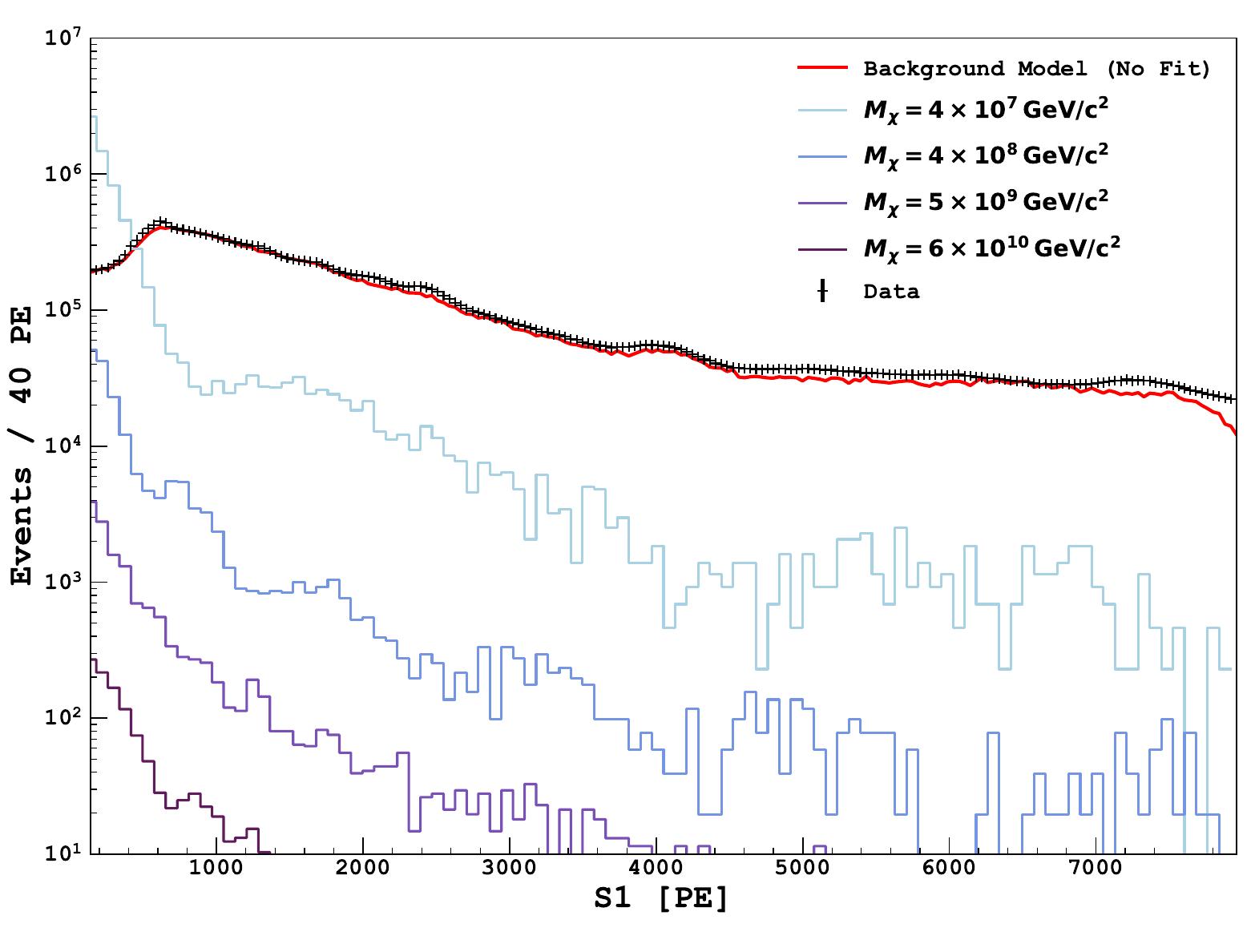}
    \setlength{\abovecaptionskip}{0pt}
    \hfill
    \caption{Energy spectrum of various UHDM models in the range [100--8000] PE. The red and black curves indicate the background model and observed energy spectrum, respectively. The different purple/blue colors show signals of different $M_{\chi}$, for $\sigma_{\chi,n}=4.3\times10^{-26} \, \mathrm{cm^2}$.}
    \label{fig:sig}
\end{figure}

The energy spectrum in the region of interest for various UHDM models are presented in figure \ref{fig:sig}. Increasing $M_{\chi}$ increases the form factor suppression, reducing the recoil energies of the UHDM passing through the TPC. Increasing $\sigma_{\chi,n}$ increases the number of energy deposits of a UHDM particle passing through the detector, shifting the signal energy spectrum towards higher energies (not shown).

\section{Results}
\label{sec:Exc}
With the signal and background models, and the collected data energy distributions, a $\mu$ (signal strength parameter) fitter is used to estimate the p-value for a particular parameter space value point ($M_{\chi}$, $\sigma_{\chi,n}$) employing the likelihood model described in reference \cite{Cowan}. Since for the case of multiply interacting UHDM, increasing (decreasing) $\mu$ ($\sigma_{\chi,n}$) will not linearly increase (decrease) the signal energy distribution, the exclusion region is reconstructed by scanning the parameter space in both $M_{\chi}$ and $\sigma_{\chi,n}$, and obtaining the p-value for $\mu=1$. If the confidence of a given parameter space point is above 90\%, the point is excluded.
Table~\ref{tab:pars} shows the nuisance parameters from radiogenic background contributions for the background-only fit; all pulls are consistent with zero ($|\frac{\theta - \theta_0}{\sigma_{\theta}}| < 0.03$), indicating that the fitted values remain close to their priors, while the significantly reduced post-fit uncertainties demonstrate that these background components are strongly constrained by the data.

\begin{figure}[h!]
    \centering
    \includegraphics[width=8.5cm]{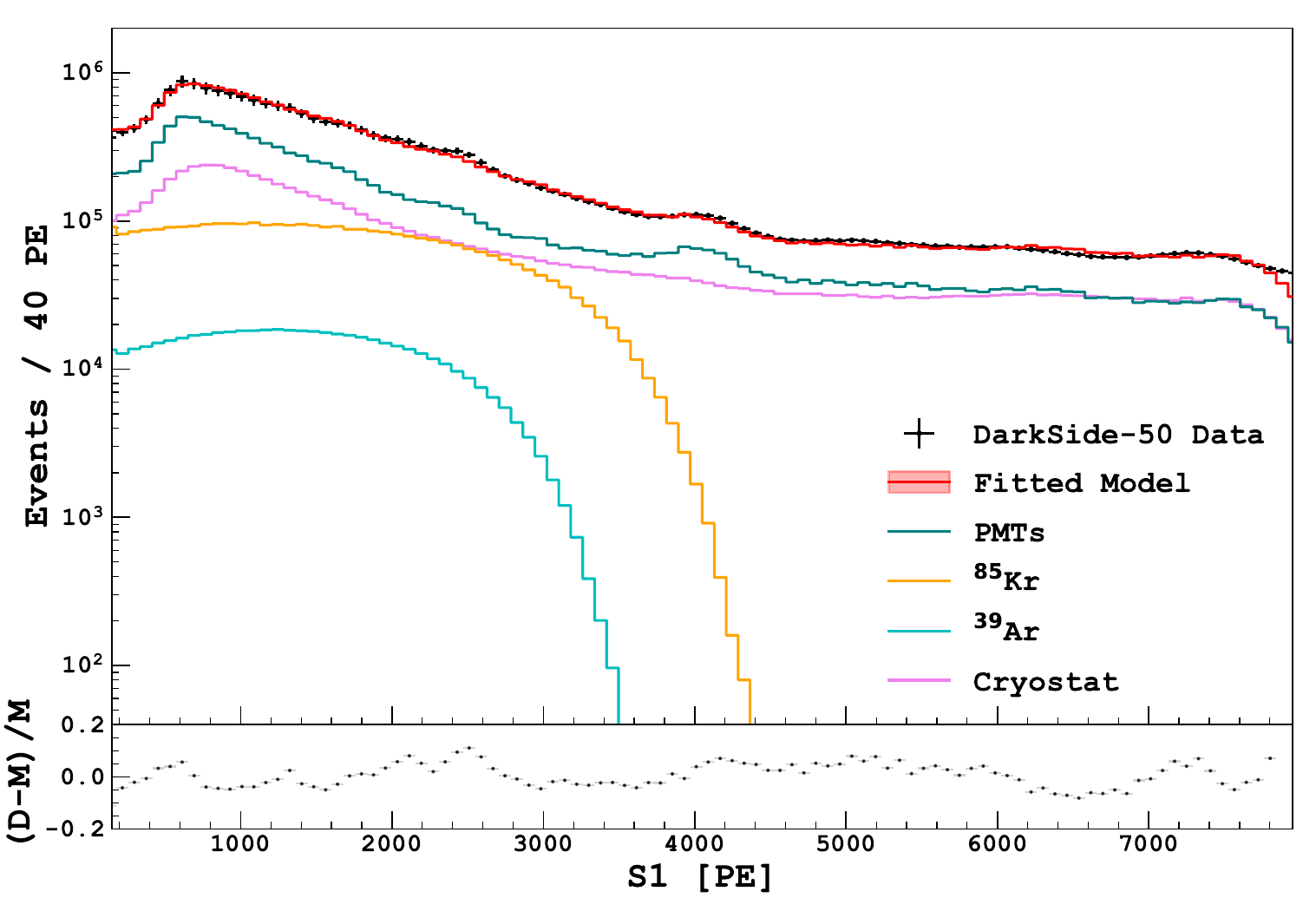}
    \setlength{\abovecaptionskip}{0pt}
    \caption{Results of the background-only fit. The bottom sub-axis shows the difference between the model and the observed data as $(\mathrm{Data}-\mathrm{Model})/\mathrm{Model}$. The light yield at 200 V/cm nominal field was $7.0\pm 0.3 \mathrm{\, PE/keV}$ \cite{DarkSideFirst}.}
    \label{fig:BkgFit}
\end{figure}

\begin{table}[htpb!]
\centering
\begin{tabular}{lcc}
\hline
\textbf{Parameter} & \textbf{Prior} [mBq/kg] & \textbf{Pull} ($\frac{\theta - \theta_0}{\sigma_{\theta}})$  \\ 
\hline
$N_{\mathrm{pmt}}$ & 259.1 & $6.70 \times 10^{-4} \pm 9.30 \times 10^{-2}$ \\ 
$N_{^{85}\mathrm{Kr}}$ & 1.86 & $-2.29 \times 10^{-2} \pm 2.01 \times 10^{-1}$ \\
$N_{\mathrm{cryo}}$ & 21.27 & $1.09 \times 10^{-2} \pm 1.87 \times 10^{-1}$ \\ 
$N_{^{39}\mathrm{Ar}}$ & 0.73 & $7.96 \times 10^{-3} \pm 1.57 \times 10^{-1}$ \\ 
\hline
\end{tabular}
\caption{Nuisance parameters for the radiogenic activities contributors in figure \ref{fig:BkgFit}. Both the priors before other scaling efficiencies and the pulls ($\frac{\theta - \theta_0}{\sigma_{\theta}}$) are shown.}
\label{tab:pars}
\end{table}

Figure \ref{fig:limit_masses2} shows the 90\% C.L. excluded region for different dark nucleon masses ($m_{\chi}$), illustrating the region of the parameter space for which the UHDM model is excluded. The bottom edge is dictated by Poisson mean value of the cross-section at which dark matter is expected to scatter in every pass through the detector, $\sigma_s$ \cite{SaturatedOverburden}. The left-most boundaries are delimited by the overburden. For lower $m_\chi$ masses, DM models fall below the sensitivity of the search. The top boundary of the exclusion curve is delimited by the geometrical cross section of the dark matter model with dark nucleon mass $m_{\chi}$. Higher cross section are un-physical and have also been ruled out by self-interaction (UHDM) limits \cite{Coskuner}. The drop-off of the top boundary of the excluded region at $M_{\chi} \gtrsim 10^9 \, \mathrm{GeV/c^2}$ is explained by the large number of energy depositions within one TPC pass, producing signals above the 8000 PE region of interest. As $m_{\chi}$ increases, the region between the geometrical cross-section and $\sigma_{s}$ becomes smaller, reducing the visible parameter space. At high masses, the recoil energies are highly reduced due to form factor effects and fall below the energy threshold of the detector. The right-hand side is delimited by the detector's exposure and dark matter flux.

We compare our results with other experiments that employ $\frac{d\sigma_{\chi,N}}{dE_R} = \frac{d\sigma_{\chi,n}}{dE_R} A^4 |F_{N}(q)|^2$, namely DEAP-3600 \cite{DEAPFirst}, XENON-1T \cite{XenonProjected}, and LZ {\cite{Lux}, and report less sensitivity at low cross sections and high masses, which is a consequence of different detector exposures and analysis approaches. The other searches performed a background-free analysis, and as previously mentioned, form factor effects begin to suppress the energy depositions resulting in very low energy deposits and a loss of sensitivity in a S1-only search. In comparison to these other experiments, we incorporate form factor effects to account for the size of the dark matter particle and extend the search for lower dark nucleon masses, down to $m_\chi \sim 10^7$ GeV/c$^2$.

\begin{figure}[htpb]
    \centering
    \includegraphics[width=8.5cm]{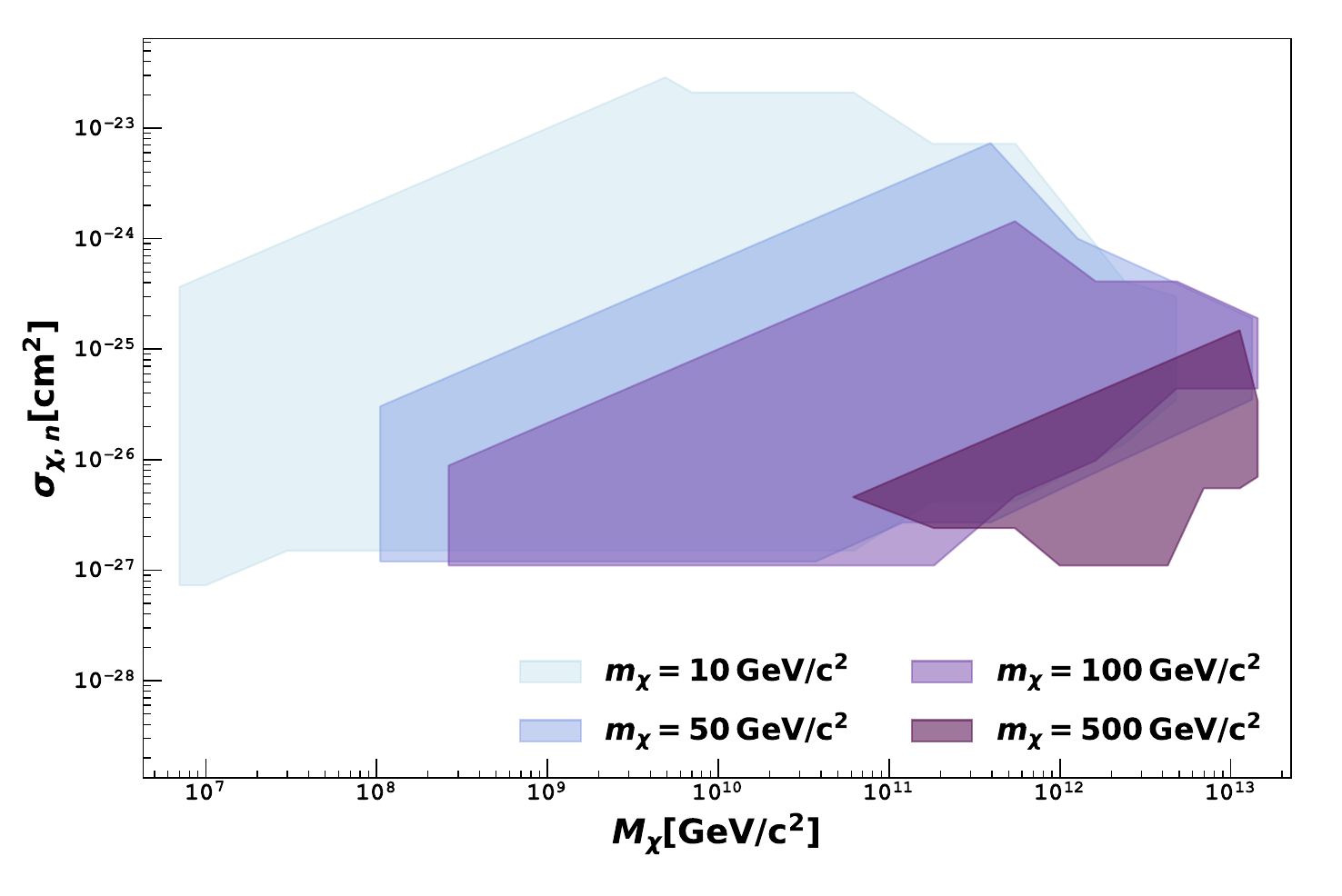}
    \setlength{\abovecaptionskip}{0pt}
    \caption{90\% C.L. excluded region for nuclear UHDM for $m_{\chi} = 10, \, 50, \, 100, \, 500 \, \mathrm{GeV/c^2}$}
    \label{fig:limit_masses2}
\end{figure}

\section{Conclusion}
\label{sec:Conc}
We present the first search for nuclear Ultra Heavy Dark Matter (UHDM) using a dual-phase liquid argon time projection chamber with the DarkSide-50 experiment. This work extends UHDM searches into a new experimental regime for dual phase TPCs. No UHDM signal events were observed, and the resulting exclusion region defines the parameter space ruled out at 90 \% C.L.
We have modeled UHDM as a composite particle and systematically varied the dark nucleon mass—an approach not previously explored in LAr TPCs. This represents an important step toward understanding the internal structure of UHDM models, as the dark nucleon mass fundamentally determines the scattering dynamics and energy deposition patterns in the detector. By deriving exclusion limits for dark nucleon masses of $m_{\chi} = 10, 50, 100, 500 \, \mathrm{GeV/c^2}$, we demonstrate how the composite nature of UHDM affects experimental sensitivity. As $m_\chi$ enters through the UHDM form factor, the resulting signal hypotheses are not directly interpolable with those of other experiments, precluding a direct numerical comparison. This work establishes a path for future UHDM searches to constrain not only the existence of ultra-heavy dark matter but also its internal composition.

\section{Acknowledgment}
\small{
The DarkSide Collaboration offers its deep gratitude to LNGS and its staff for their invaluable technical and logistical support.
The authors also thank the Fermilab Particle Physics, Scientific, and Core Computing Divisions.
 Construction and operation of the DarkSide-50 detector was supported by the U.S. National Science Foundation (NSF) (Grants No. PHY-0919363, No. PHY-1004072, No. PHY-1004054, No. PHY-1242585, No. PHY-1314483, No. PHY-1314501, No. PHY-1314507, No. PHY-1352795, No. PHY-1622415, and associated collaborative Grants No. PHY-1211308 and No. PHY-1455351),
 the Italian Istituto Nazionale di Fisica Nucleare,
 the U.S. Department of Energy (Contracts No. DE-FG02-91ER40671, No. DEAC02-07CH11359, and No. DE-AC05-76RL01830), the Polish NCN (Grant No. UMO-2023/51/B/ST2/02099)
 and the Polish Ministry for Education and Science (Grant No. 6811/IA/SP/2018).
 We also acknowledge financial support from the French Institut National de Physique Nucléaire et de Physique des Particules (IN2P3), the IN2P3-COPIN consortium (Grant No. 20-152), and the UnivEarthS LabEx program (Grants No. ANR-10-LABX-0023 and No. ANR-18-IDEX-0001),
 from the S{\~a}o Paulo Research Foundation (FAPESP) (Grant No. 2017/26238-4 and 2021/11489-7),
 from the Interdisciplinary Scientific and Educational School of Moscow University “Fundamental and Applied Space Research,”
 from the Program of the Ministry of Education and Science of the Russian Federation for higher education establishments, Project No. FZWG-2020-0032 (2019-1569),
 the International Research Agenda Programme AstroCeNT (MAB/2018/7) funded by the Foundation for Polish Science from the European Regional Development Fund,
 and the European Union’s Horizon 2020 research and innovation program under Grant Agreement No. 952480 (DarkWave), the National Science Centre, Poland (2021/42/E/ST2/00331),
 and from the Science and Technology Facilities Council, United Kingdom.
  I. Albuquerque was partially supported by the Brazilian Research Council (CNP).
}

\bibliography{References} 

\end{document}